\documentstyle[epsfig,prl,aps]{revtex}

\twocolumn

\begin{document}

\title{Hard Spheres in Vesicles: Curvature-Induced Forces and
Particle-Induced Curvature} \author{A. D. Dinsmore, D. T. Wong, Philip
Nelson, A. G. Yodh} \address{Department of Physics and
Astronomy\\University of Pennsylvania\\
Philadelphia, PA 19104 U.S.A.} \date{\today} \maketitle


\begin{abstract}
We explore the interplay of membrane curvature and nonspecific binding
due to excluded-volume effects among colloidal particles inside lipid
bilayer vesicles.  We trapped submicron spheres of two different sizes
inside a pear-shaped, multilamellar vesicle and found the larger
spheres to be pinned to the vesicle's surface and pushed in the
direction of increasing curvature.  A simple model predicts that hard
spheres can induce shape changes in flexible vesicles.  The results
demonstrate an important relationship between the shape of a vesicle
or pore and the arrangement of particles within it.
\end{abstract}

\vspace{.3in}

Entropic excluded-volume (depletion) effects are well known to lead to
phase separation in the bulk of colloids and emulsions consisting of
large and small particles with short-range repulsive interactions
\cite{DINSPRE,DHONT,STEINER,BIBETTE,BARTLETT,ILETT}.  More recently,
attraction of the large particles to flat, hard walls
\cite{KAPLAN,2KAPLAN} and repulsion from step edges \cite{DINSNAT}
have been demonstrated in binary hard-sphere mixtures.  A key concept
suggested in these papers is that the geometric features of the
surface can create ``entropic force fields'' that trap, repel or
induce drift of the larger particles.  This mechanism is not limited
to suspensions of micron-sized particles; it may play a role in ``lock
and key'' steric interactions on smaller macromolecular length scales.
For example, the shape of pores and liposomes inside cells is likely
to affect the behavior of macromolecules confined within them
\cite{MINTON}.

In this Letter, we present experimental results that demonstrate new
entropic effects at surfaces.  In particular, the behavior of
particles confined within vesicles reveals quantitatively the striking
effect of membrane curvature.  We first discuss experiments probing
the behavior of a microscopic sphere trapped inside a rigid,
phospholipid vesicle.  Adding much smaller spheres to the mixture
changes the distribution of the larger sphere in a way that depends on
the curvature of the vesicle wall (see Fig.~\ref{fig:vesicle}(b) and
(c)).  The results are consistent with the depletion force theory and
illustrate a new mechanism for the size-dependent arrangement of
particles within pores.  We then explore theoretically some
consequences of replacing the rigid wall with a flexible one.  The
entropic curvature effects can overcome the membrane's stiffness,
leading to a new mechanism for shape changes in vesicles.

We first briefly review depletion effects in mixtures of microscopic
hard spheres of two different sizes.  Moving two of the larger spheres
toward one another does not change their interaction energy (which is
zero for hard spheres) but does increase the volume accessible to the
other particles (Fig.~\ref{fig:depletion}).  The resulting gain in
entropy reduces the free energy of the system by $(3/2) \alpha
\phi_{S} k_{{\rm B}}T$ \cite{ASAK,VRIJ}.  Here, $\alpha$ is the ratio
of large to small radii ($R_{L}/R_{S}$), $\phi_{S}$ is the
small-sphere volume fraction, and $k_{{\rm B}}T$ is Boltzmann's
constant times the absolute temperature.  This simple result relies on
the approximation that the small spheres are a structureless ideal gas
and that the large-sphere volume fraction, $\phi_{L}$, is small.  The
reduction of free energy produces an ``entropic force'' that pushes
the large spheres together.  When the large sphere is moved to a flat
wall, moreover, the overlap volume and the free-energy loss are
approximately doubled \cite{KAPLAN}.  In binary hard-sphere mixtures,
these effects are known to drive crystallization of large spheres in
the bulk \cite{DINSPRE,DHONT,BARTLETT} and at a flat surface
\cite{DINSPRE,KAPLAN,DINSEL,POON}.  Furthermore, the shape of the wall
can lead to entropic forces in a specific direction {\em along the
wall}.  For example, the larger spheres are locally repelled from an
edge cut into the wall \cite{DINSNAT} and attracted to a corner ({\it
i.e.}  where the ``wall'' meets the ``floor'') \cite{DINSCORNER}. If
the wall has constantly-changing radius of curvature, these forces are
predicted to act everywhere along it \cite{DINSNAT}.  As shown in
Fig.~\ref{fig:depletion} (c), when the large sphere is near the wall,
the overlap volume depends on the wall's curvature radius.  The large
sphere will therefore move in the direction of increasing curvature to
minimize the small spheres' excluded volume.

To measure this surface entropic force, we have studied the
distribution and dynamics of microscopic nearly-hard spheres trapped
inside rigid, pear-shaped vesicles.  The vesicles were prepared from a
phospholipid (1-Stearoyl-2-Oleoyl-sn-Glycero-3-Phosphocholine (SOPC),
Avanti Inc., U.S.A.), dissolved in chloroform (25~mg/mL).  After
evaporating the chloroform from 200~$\mu$L of SOPC solution, we added
100~$\mu$L of salt water with charge-stabilized polystyrene spheres
(Seradyn, IN, USA) in suspension.  The salt (0.01~M NaCl) served to
screen out electrostatic forces over a distance of $\approx 5$~nm, the
Debye-Huckel screening length \cite{KAPLAN}.  Thus, our sample closely
resembled an ideal hard-sphere and hard-wall (HSHW) system.  Rigid,
multilamellar vesicles of diverse shapes and sizes immediately formed
with colloidal spheres trapped inside.  We injected the solution into
a 10-$\mu$m thick glass container for viewing under an optical
microscope (100x objective with 1.30 numerical aperture, in
transmission mode).  Images of a planar slice, 600~nm thick, through
the center of vesicles were captured, then later digitized
(Fig.~\ref{fig:vesicle}(a)).

We quantified the behavior of the mixture by measuring its free
energy, $F({\bf r})$, as a function of the position of a sphere of
radius $R_L=0.237\,\mu$m.  First, using NIH Image software, we
determined the in-plane position of the sphere's center of mass, ${\bf
r}$, ($\pm 0.08 \mu$m) when it appeared in the imaged slice.  From
these data, we used two techniques to extract $F({\bf r})$.  In the
first, the number of times $N({\bf r}_{i})$ the sphere appeared at a
given bin located at ${\bf r}_{i}$ defined the density distribution.
Assuming that each measurement event was independent of the others,
$N({\bf r}_{i})$ follows the Boltzmann distribution: $N({\bf r}_{i})
\propto {\rm exp}(-F({\bf r}_{i}) /k_{{\rm B}}T)$.  Since a systematic
error can arise if events are not completely independent ({\it i.e.}\
not separated by an infinite time), we waited a minimum of 0.6~s
between measurements.  During this time the mean square displacement
was $\approx$ 0.3~$\mu$m, larger than the 0.07-0.13~$\mu$m bin sizes.
We also collected data over a period of 30-80 minutes, enough time for
the sphere to explore all of the available space.  We therefore used
the logarithm of the sphere's distribution to extract the free energy.

The dynamics of the diffusing 0.237-$\mu$m sphere provided the second
way to measure $F({\bf r}_{i})$, as described in \cite{CROCKER}.  The
region along the inner surface of the vesicle was divided into several
equal-area bins.  We considered only events in which the center of the
large sphere was within $0.28~\mu$m of contact with the surface in
consecutive frames (separated by a time $\tau$).  From our videotape,
we counted the number of times ($N_{ij}(\tau)$) the sphere was located
in bin $j$ at time $t$ and in bin $i$ at time $t + \tau$.  The
transition probability matrix, $P_{ij}(\tau)$, is given by
$N_{ij}(\tau)/N({\bf r}_{j})$.  The measured matrix $P_{ij}$ contains
information about the equilibrium state of the system.  In particular,
the eigenvector ($\hat{{\bf e}}$) of $P_{ij}$ with unit eigenvalue is
proportional to the Boltzmann factor: $\hat{{\bf e}}_{i} \propto {\rm
exp}(-F({\bf r}_{i})/k_{{\rm B}}T)$.  We obtained consistent results
with $\tau$ = 0.1, 0.2 and 0.3~s and with various bin sizes.  This
technique avoids fit parameters and possible systematic errors arising
from the density-distribution approach.  It also avoids potential
errors arising from a slow change in shape of the vesicle, which would
affect the distribution averaged over a long time but would not affect
the short-time particle dynamics.
We report here the results from two different samples. The first
``control'' sample contained a solitary 0.237-$\mu$m sphere (no small
spheres) diffusing freely inside a vesicle.  The measured density
distribution is shown in Fig.~\ref{fig:vesicle}(b).  The sphere
distribution is uniform: there is no significant interaction between
the vesicle wall and the polystyrene sphere, as expected due to the
very short Debye-Huckel screening length.

The behavior changed noticeably when small spheres were added to the
interior of a vesicle.  In Fig.~\ref{fig:vesicle}(c), we show the
distribution of the $R_L=0.237\,\mu$m sphere in a binary mixture with
$\phi_{S} = 0.3, R_{S} = 0.042 \mu$m ($\alpha = 5.7$).  The
distribution is highly non-uniform, with a significantly higher
probability of finding the large sphere within about $0.28~\mu$m of
the surface.  The apparent width of this ``surface'' region exceeds
$2R_{S}$ due to uncertainties in the particle positions and, possibly,
due to a slight tilt of the vesicle wall away from vertical.


We measured the average number of times the large sphere appeared in
each bin within $0.28~\mu$m of the surface and the average number per
bin in the bulk.  We defined the natural logarithm of the ratio of
these numbers as $F_{0}/(k_{B}T)$, a measure of the average strength
of the depletion attraction.  We found $F_{0} = (2.2 \pm 0.5) k_{B}T$.


Theoretically, $F_0=\ln\,\int_{R_L}^{R_L+0.28~\mu{\rm m}}[{\rm
d}r/0.28~\mu{\rm m}]\,{\rm e}^{-V(r)/k_{B}T}$, where $r$ is the
distance from the large-sphere center to the wall and $V(r)$ is given
by the depletion force model.  Although the vesicle wall was curved,
we can predict a lower bound of $F_{0}$ by assuming the wall is flat:
\begin{equation}
V(r) = \frac{-k_{B}T \phi_{S}}{4 R_{S}^{3}} (r - R_{L} - 2 R_{S})^2 
(r + 2R_{L}+R_{S}).
\end{equation}
Putting in the numbers gives a theoretical prediction of $F_0=1.97
k_{B}T$.  Here we have neglected thermal fluctuations in the shape of
the vesicle wall and residual electrostatic repulsion.  Constraining
the large sphere to lie 5~nm away from the wall, for example, reduces
the predicted result to $1.45~k_{B}T$.  The $0.7$-$k_{B}T$ difference
from our measured result is likely due to the curvature of the wall,
which would enhance the observed $F_{0}$ as discussed in the following
paragraph.  These results are consistent with recent calculations
showing that the ideal-gas approximation accurately predicts the
depletion well depth at contact (although it misses the relatively
weak, long-range depletion repulsion) \cite{DICKMAN}.


The distribution in Fig.~\ref{fig:vesicle}(c) demonstrates a higher
large-sphere occupation where the vesicle surface is more curved.
>From the sphere dynamics, we obtained the total free energy, $F$, as a
function of sphere position when it was near the surface.  The inset
of Fig.~\ref{fig:F} shows the measured $F$ as a function of the
position, $s$, along the perimeter.  The slope of this curve reveals a
maximum force of $20 {\rm x} 10^{-15}N$ pushing the large sphere along
the wall.  We also determined the curvature radius at points along the
wall using our images of the vesicle.  In Fig.~\ref{fig:F}, we plot
the free energy as a function of wall curvature radius, $R_{C}$.  The
free energy decreased by approximately $1.5 k_{{\rm B}}T$ when the
curvature radius decreased from 20~$\mu$m to 2~$\mu$m.  For comparison
to our results, we used the simple depletion-force model described
above (Fig.~\ref{fig:depletion}(c)) to calculate the free energy as a
function of wall curvature (assuming a locally spherical wall):
$F(R_{C}) = -P_{S} V_{{\rm overlap}}(R_{C})$.  Here, $P_{S}$ is the
small-particle osmotic pressure (from the ideal-gas law) and $V_{ {\rm
overlap}}(R_{C})$ is the size of the excluded-volume overlap.  The
large sphere was assumed to touch the vesicle wall.  The theory,
represented by the solid line in Fig.~\ref{fig:F}, is consistent with
the experimental results.  For both plots, the free energy was set to
0 in the limit of large curvature radius (flat wall).  Since we could
only measure $R_{C}$ in the image plane, we assumed that this also
determined the out-of-plane curvature.  This approximation probably
explains the scatter in our data points.

So far we have shown that curvature of a {\it rigid} wall induces
forces on the large spheres in binary particle mixtures.  To conclude
this Letter, we consider theoretically a binary hard-sphere mixture in
the presence of {\em flexible} walls.  This extra degree of freedom
introduces a competition between depletion and curvature energy that
can produce a variety of new phenomena.  For example, we expect that
under some circumstances, the membrane will spontaneously bend around
the large sphere (Fig.~\ref{fig:wrap}).  We anticipate that these
effects will be observable and that understanding this mechanism will
be essential to predict the behavior of unilamellar phospholipid
vesicles containing particles of different sizes.

The membrane will spontaneously envelop the large sphere only if the
resulting total free energy change, $\Delta F$, is negative.  Our
experiment indicates that the membrane-particle interactions are very
short-ranged, so we use the hard-sphere and hard-wall approximations.
We therefore divide the full free energy into two terms, one
representing the net adhesion induced by the colloidal spheres, the
other the vesicle's bending energy: $F = F_{{\rm adh}} + F_{{\rm
ves}}$.


The experiment described above shows that for rigid membranes,
$F_{{\rm adh}}$ can be well described by the depletion-force
model. For flexible membranes we must add a new contribution to
$F_{{\rm adh}}$ to account for the ``steric repulsion'' effect
\cite{HELF_FLUCT,SAFINYA}. For $R_S\ll R_L$ we may approximate the
latter by the interaction between a flat hard wall and a fluctuating
membrane at constant mean separation $x$.  This free energy equals
$0.06(k_{{\rm B}}T)^{2}/\kappa x^2$ per unit
area~\cite{LZ89,SEIFERT2}. For unilamellar phospholipid membranes, the
bending stiffness $\kappa \approx 15k_{{\rm B}}T$ \cite{SDM}.
Approximating the change of excluded volume upon adhesion by $2R_S-x$
per unit area and again taking the small-sphere volume fraction to be
$\phi_S=0.3$, we minimize the total adhesive energy (depletion
attraction plus steric repulsion) to find $x\sim 0.5 R_S$ and the
adhesive energy $F_{{\rm adh}}=-0.091k_{{\rm B}}T\cdot A^*/{R_ S}^2$,
where $A^{*}$ is the area of contact between the membrane and the
sphere (Fig.~\ref{fig:wrap}(b)) \cite{FNPN}.

The vesicle's elastic energy, $F_{{\rm ves}}$, depends only on the
membrane curvature.  We neglect the constant-volume and constant-area
constraints of a vesicle and instead consider an infinite membrane.
Thus the local curvature energy change upon adhesion is just $F_{{\rm
ves}}=2\kappa A^*/{R_L}^2$ \cite{CANHAM,HELFRICH}. Combining $F_{{\rm
adh}}+F_{{\rm ves}}$, we see that adhesion and engulfment are favored
when $\alpha\equiv R_L/R_S>\sqrt{2\kappa/0.091k_{{\rm B}}T }=18$.  A
more detailed analysis of the deformation of vesicles at a generically
sticky surface, specialized to the case of interest here, gives a
similar result \cite{SEIFERT1}. The required regime for $\alpha$ is
easily accessible, using either polymers or spheres as the small
particles.

In conclusion, we have measured for the first time a curvature-induced
entropic force in a system of hard spheres trapped in a rigid vesicle.
The results show that the distribution of particles within the vesicle
is strongly affected by the local shape of the vesicle wall.
Furthermore, a simple estimate predicts shape transitions of
unilamellar phospholipid vesicles induced by particles inside the
vesicle.  The ideas presented here suggest a way to understand several
phenomena in cellular interiors and complex fluids inside porous
media.  For example, in a vesicle whose membrane contains multiple
species of lipids, the lipids can segregate into regions with
different curvatures \cite{LIPOWSKY}.  In such a sample, the particle
distribution, which depends on curvature, would correlate with the
local {\em composition} of the membrane. This mechanism may provide
considerable specificity in associations between macromolecules and
lipids.  Finally, when the size ratio is more extreme or the vesicle
wall is less rigid, the spheres may be able to induce budding and
fission of the vesicle.

\begin{acknowledgements}

We are happy to thank D.  J. Pine, T. C.  Lubensky, U. Seifert,
D. A. Weitz and J. C. Crocker for their helpful discussions.  AGY was
supported in part by NSF grant DMR96-23441 and by the NSF-PENN MRSEC
Program.  PN was supported in part by NSF grant DMR95--07366, and by
US/Israeli Binational Foundation grant 94--00190.  DTW acknowledges
support from the Howard Hughes Medical Institute.
\end{acknowledgements}

\begin{onecolumn}

\begin{figure}[h]
\epsfig{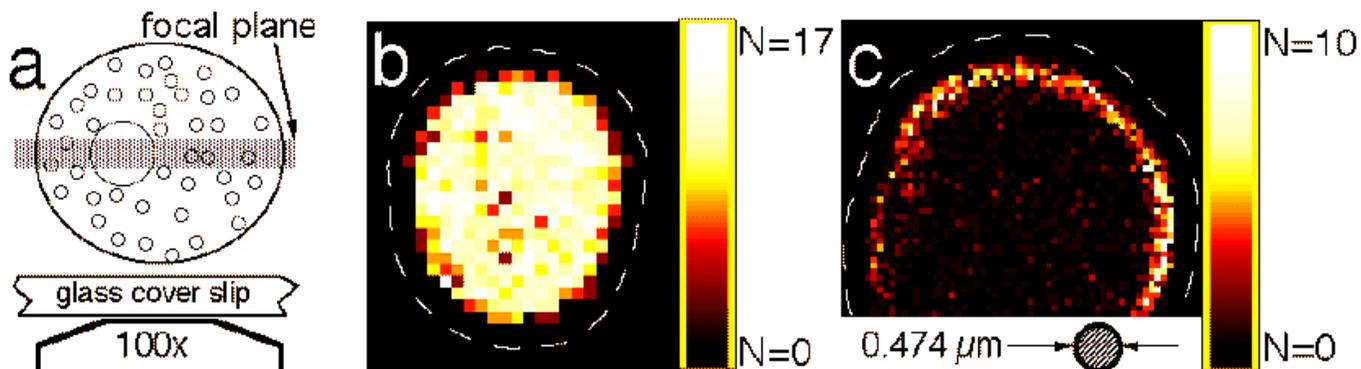}
\caption{(a) Cartoon of the 600-nm thick slice through an SOPC vesicle
imaged with an optical microscope.  We measured the in-plane positions
of the larger colloidal sphere when it was in focus.  (b)(color)
Probability distribution of a single 0.237-$\mu$m-radius polystyrene
sphere inside a vesicle (no small spheres).  The white dashed line is
the edge of the vesicle, and the colored points indicate the number of
times, $N$, the center of the sphere was observed in a bin located at
a given point.  There were 2000 events and the bins were 130~nm x
130~nm.  The sphere simply diffused freely throughout all of the
available space.  (c) (color) Same as in (b), but with a vesicle that
also contained small spheres $(\phi_{S} = 0.30, R_{S}=0.042~\mu$m$,
\alpha = 5.7)$.  There were 2300 events and the bin size was 65~nm.
The large sphere was clearly attracted to the vesicle wall, especially
where the vesicle was most curved.}
\label{fig:vesicle}
\end{figure}

\end{onecolumn}
\begin{twocolumn}

\begin{figure}[h]
\epsfig{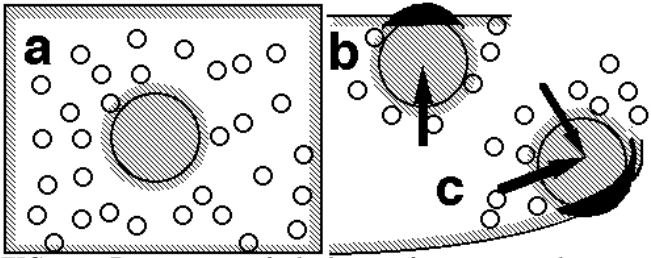}
\caption{Description of depletion forces in a binary hard-sphere
mixture.  The centers of mass of the small spheres are excluded from
the hatched regions, within one radius of the surface of the large
sphere and of the walls.  In (a), the volume accessible to the
small-sphere centers, $V_{{\rm acc}}$ is the total volume minus the
hatched regions.  (b) When the large sphere moves to the wall, the
excluded regions overlap, as shown in black, and $V_{{\rm acc}}$
increases by this amount.  The small spheres' entropy therefore
increases.  (c) Because of the changing wall curvature, the size of
the overlap depends on sphere position.  The large sphere will move
along the wall to maximize the size of the overlap region, as
indicated by the arrow.}
\label{fig:depletion}
\end{figure}

\begin{figure}
\epsfig{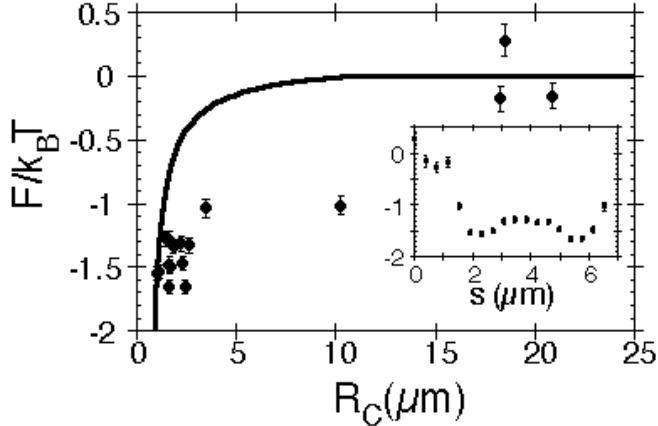}
\caption{The total free energy of the binary mixture, $F$, when the
large (0.237-$\mu$m) sphere was at the vesicle wall.  $F$ is plotted
in units of $k_{{\rm B}}T$ as a function of the wall curvature radius,
$R_{C}$.  The symbols represent measurements and the line represents
the results of the calculation described in the text.  (Inset)
$F/k_{B}T$ {\it vs.} position, $s$, along the perimeter of the
vesicle.  The origin ($s$=0) is located at the lower left of the
vesicle shown in Fig.~1(c).}
\label{fig:F}
\end{figure}

\begin{figure}[h]
\epsfig{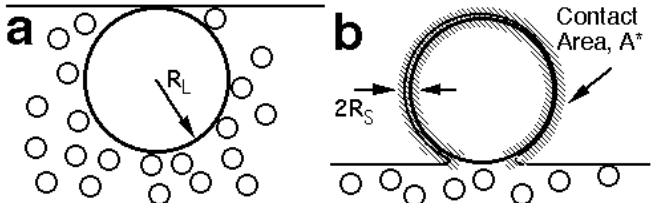}
\caption{Cartoon demonstrating how hard spheres can induce a change in
shape of a vesicle.  (a) When the membrane is flat, the
excluded-volume overlap is minimized and the curvature energy is zero.
(b) When the vesicle envelops the large sphere, the curvature energy
increases, but so does the excluded-volume overlap (hatched region).}
\label{fig:wrap}
\end{figure}

\end{twocolumn}

\begin{references}
\bibitem{FNPN} For extremely flexible membranes with $\kappa\sim
k_{{\rm B}}T$ the adhesion is lost completely. Details of this
calculation will appear elsewhere.
\end{references}
\end{document}